\documentclass[aps,prb,reprint,groupedaddress]{revtex4-1}
\usepackage{graphicx}
\usepackage{color}
\renewcommand\Im{\mbox{Im}}

\begin{document}

\preprint{v1.0}

\title{Dynamics of optical excitations in a Fe/MgO(001) heterostructure from
time-dependent density functional theory}

\author{Markus Ernst Gruner}
\email[]{Markus.Gruner@uni-due.de}
%\affiliation{University of Duisburg-Essen and Center for Nanointegration, CENIDE, 47048 Duisburg, Germany}
\affiliation{Faculty of Physics and Center for Nanointegration, CENIDE, University of Duisburg-Essen, 47048 Duisburg, Germany}
\author{Rossitza Pentcheva}
\email[]{Rossitza.Pentcheva@uni-due.de}
\affiliation{Faculty of Physics and Center for Nanointegration, CENIDE, University of Duisburg-Essen, 47048 Duisburg, Germany}

\date{\today}

\begin{abstract}
In the framework of real-time time-dependent
density functional theory (RT-TDDFT)
% Revision B.3a
we unravel the layer-resolved dynamics of excited carriers in a
(Fe)$_1$/(MgO)$_3$(001) multilayer
after an optical excitation with a frequency below
the band gap of bulk MgO.
Substantial transient changes to the electronic structure, which persist after the
duration of the pulse, are mainly observed for in-plane polarized
electric fields, corresponding to a laser pulse arriving perpendicular to the
interface.
While the strongest charge redistribution takes place in the Fe layer,
a time-dependent change in the occupation numbers is visible in all layers,
mediated by the presence of interface states.
The time evolution of the layer-resolved time-dependent occupation numbers
indicates a strong orbital dependence with the depletion from in-plane orbitals
(e.\,g.,  $d_{x^2-y^2}$ of Fe) and accumulation in  out-of-plane orbitals ($d_{3z^2-r^2}$ of Fe and $p_z$ of apical oxygen).
We also observe a small net charge transfer away
from oxygen towards the Mg sites even for MgO layers which are not
directly in contact with the metallic Fe.
\end{abstract}

% \keywords{}

\maketitle

\section{Introduction}\label{sec:intro}
Resolving the impact of optical excitations on the different degrees of freedom present
in solid state materials  on ultrashort time scales
provides important insight
in the coupling mechanisms
present in modern functional materials.
% Revision: B.1
This is relevant for the preparation of novel transient states of matter, which cannot be
reached in an ergodic process under equilibrium conditions, but
can be accessed
through the excitation with a strong laser
pulse.\cite{cn:Ernstorfer09,cn:Ichikawa11,cn:Stojchevska14,cn:Huebener17}
Such questions are
addressed in pump-probe experiments
using optical or X-ray pulses with femto- or picosecond delay.
A large fraction of these experiments are %also
devoted to resolving
and controlling the dynamics of magnetization reversal processes\cite{cn:Stamm07,cn:Radu11}
(for an overview, see also Ref.\ \onlinecite{cn:Eschenlohr18}
and references therein)
and the particular dynamics of spin and orbital moments in metals.\cite{cn:Boeglin10} 
Furthermore, the spin-relaxation time of hot carriers in Au was studied in an optical pump
second optical harmonic probe experiment  on a Au/Fe/MgO(001) multilayer stack.\cite{cn:Melnikov11}
Another outstanding example is the investigation of the dynamics of the electronic structure
in the laser-induced insulator-to-metal transition in VO$_2$.\cite{cn:Cavalleri05}

From the theoretical point of view, time-dependent density functional theory (TDDFT)
has evolved to a major workhorse in describing the excitation of
molecules and extended systems by laser pulses
(for recent reviews, see, e.g., Refs.\ \onlinecite{cn:Botti07,cn:Marques12,cn:Sharma14,cn:Olevano18}).
For weak laser pulses, linear-response TDDFT (LR-TDDFT) with an appropriately
chosen exchange correlation-kernel can provide an adequate description of the
frequency dependent excitation spectrum.\cite{cn:Maitra16}
For strong pulses, where nonlinear effects become relevant,
real-time TDDFT (RT-TDDFT) approaches based on the explicit
solution of the time-dependent Kohn-Sham-equations with an external time-dependent
electrical potential need to be employed.
The high computational demand limits the system size, but increasing computing power and
ongoing code-development help to overcome the limitations. 
A large fraction of TDDFT calculations is concerned with
zero-dimensional systems, as molecules or clusters (e.g., Refs.\ \onlinecite{cn:Burke05,cn:Takimoto07,cn:Lopata11,cn:Cocchi14}).
More recently, also bulk systems are addressed with RT-TDDFT,
for instance, to disentangle the demagnetization processes in
metals\cite{cn:Krieger15,cn:Elliott16,cn:Krieger17}
or to describe light-matter-interaction
% Revision: B.1
in semiconductors,\cite{cn:Sato14b,cn:Tancogne17,cn:Pemmaraju18} including the
numerical simulation of pump-probe experiments in Si \cite{cn:Sato14a}.
Increasingly,  also time-dependent processes in complex systems were investigated, such as
reactions at surfaces,\cite{cn:Grotemeyer13,cn:Miyamoto17}
magnetic transitions in metal multilayers\cite{cn:Dewhurst18,cn:Chen19}
or ultrafast charge-transfer dynamics in van-der-Waals coupled
dichalcogenide 
layers.\cite{cn:Wang16,cn:Ji17,cn:Buades18}

% Revision: B.1
Interfaces are ubiquitous in many of the above materials, as these are grown on substrates or designed as
multilayer systems. These may decisively influence the dynamics and dissipation of optical excitations.
Therefore the
spatial propagation of excited carriers through an interface can be considered as an
important fundamental research question.
In the present work, we will concentrate on a system consisting of a metallic part and a wide
band-gap semiconductor or insulator.
We have selected the paradigmatic system Fe/MgO(001), for which
the electronic and transport properties have been intensively
studied in the
past\cite{cn:Butler01,cn:Mathon01,cn:Tiusan04,cn:Waldron06,cn:Belashchenko05,cn:Heiliger08,cn:Peralta08,cn:Rungger09,cn:Feng09,cn:Abedi10,cn:Raza11}
due to its relevance for spintronics applications as a tunneling magnetoresistive element in hard disk read
heads\cite{cn:Parkin04,cn:Yuasa04} or spin diode or rectifier for
magnetic logic elements.\cite{cn:Iovan08}
The choice of the system was also motivated by recent
optical-pump, X-ray-probe experiments, where
the optical pump pulse was designed to excite the metallic subsystem, only.\cite{cn:Rothenbach19}

The purpose of the present work is 
to explore with RT-TDDFT under which conditions an optically induced electronic
excitation may propagate into and possibly through the interface.
Due to the high computational cost of this method, we concentrate here
on a minimal heterostructure containing a single Fe layer and three layers of MgO,
i.\,e., Fe$_1$/(MgO)$_3$(001).
We consider laser pulses with a frequency, which is lower than the band gap of MgO, but sufficiently large
to excite electrons in Fe to a level close to or above
the conduction band minimum of MgO.
Our main focus is then on the temporal and spacial evolution
of electronic density and orbital polarization as  a function of the laser
pulse polarization direction and frequency.

The paper is structured as follows:
After a presentation of the computational details
in Sec.\ \ref{sec:compute}, a brief discussion of the ground state geometry and electronic structure
of the Fe$_1$/(MgO)$_3$(001) heterostructure is given in
Sec.\ \ref{sec:static}.
% Revision A.2
In Sec.\ \ref{sec:LRTDDFT},
we discuss the strongly anisotropic frequency dependent properties of the system, which are related to
the imaginary part of the dielectric tensor obtained with LR-TDDFT.
Sec.\ \ref{sec:dynamics} reports the real-time evolution of the electronic
system following an excitation with optical pulses as a function of frequency and
polarization direction.
Finally, the results are summarized in Sec.\ \ref{sec:conclusion}.

\section{Computational details}
\label{sec:compute}
The ground state properties (lattice parameters and atomic positions) were
obtained with
the VASP plane wave code\cite{cn:VASP1} (version 5.4.4)
using the generalized gradient approximation by
Perdew, Burke and Ernzerhof (PBE) for exchange and correlation\cite{cn:Perdew96}
with a plane wave cutoff
of 500\,eV. We employed potentials with the valence
configuration $2s^22p^4$ for O, $3s^2$ for Mg and $3d^74s^1$ for Fe, constructed for use with
the projector augmented wave method (PAW).\cite{cn:VASP2}
Brillouin zone integration was carried out on a $12\times 12\times 4$ k-mesh
with Gaussian-type Fermi-surface
smearing ($\sigma$$\,=\,$0.1\,eV).
Convergence criteria for the electronic selfconsistency cycle and for internal position and lattice parameters 
are $10^{-7}\,$eV  and $10^{-5}\,$eV, respectively.
Since we are not primarily interested in magnetization dynamics,
spin-orbit coupling was not considered in our calculations. % , which allows us

The electronic structure, optical absorption spectra and time-evolution after optical excitation
were calculated with the previously optimized geometry
using  the Elk full-potential augmented plane wave code.\cite{cn:Elk436} For consistency with the real-time calculations we have selected the 
local density approximation (LDA) for the exchange-correlation functional of
Perdew and Wang (PW92).\cite{cn:Perdew92}
% Revision: A2
Site-resolved density of states and magnetic moments
are in close agreement with the VASP
calculations using PBE.
A cutoff parameter $R\,K_{\rm max}$$\,=\,$7 for the plane waves  and a maximum angular momentum
$l_{\rm max}$$\,=\,$7 for the APW functions was used 
in combination with  Fermi-type smearing, corresponding
to an electronic temperature of $T$$\,=\,$316\,K and muffin tin radii of
1.139\,\AA{} for Fe, 1.164\,\AA{} for Mg and 0.855\,\AA{} for O.
Convergence criterion for electronic selfconsistency was a root mean square change
of $10^{-7}$ in the Kohn-Sham potential.

The dynamic evolution of an optical excitation was calculated with TDDFT
in the real-time (RT) domain, using adiabatic LDA (ALDA), which is local in space and time. % The ionic positions were kept fixed.
We used a $8\times 8\times 3$ mesh in reciprocal space and the above mentioned technical parameters.
% Revision: A2
The time propagation was carried out for at least 42\,fs,
discretized with a timestep of
0.1\,a.u.$\,=\,$$2.419\,\times\,10^{-18}\,$s.
The numerical stability of the
time evolution was tested in an additional run without an externally applied laser pulse,
confirming that the system remains in its initial state.
The laser pulse was modeled by a time-dependent but spatially constant vector potential
$\vec{A}(t)$, which contributes to the kinetic energy in the time-dependent
Kohn-Sham equations
in terms of the
generalized momentum operator $\hat{\vec{p}}=-\frac{i\,\hbar}{m_e}\,\nabla+\frac{e_0}{c}\,\vec{A}(t)$.
We simulated laser pulses with different frequencies. In all cases, the waves
were convoluted with a Gaussian with full-width at half-maximum of 5.81\,fs,
which corresponds to an effective peak time of 11.6\,fs.
Finally, we scaled the amplitudes to obtain a constant peak power density
of 4.2\,TW/cm$^2$ for all pulses. Multiplied with the full-width at half-maximum, this
yields a laser fluence of 25\,mJ/cm$^2$, which is a typical magnitude in pump-probe
experiments.\cite{cn:Bierbrauer17,cn:Eschenlohr17}

% Revision B.3a
The dynamic changes of the occupation numbers
during and after
applying a laser pulse were monitored by the time-dependent spectral function
$D(E,t)$, see Ref.\ \onlinecite{cn:Dewhurst18}.
% Revision: A3
The spin-resolved $D_{\sigma}(E,t)$ is calculated from the projection of the
time-propagated orbitals $\Phi(\vec{r},t)_{j\vec{k}\sigma}$ onto the ground-state Kohn-Sham
orbitals at $t=0$,
which can be used to define time-dependent and spin-resolved occupation
numbers (with spin-index $\sigma$): 
\begin{equation}
  g_{i\vec{k}\sigma}(t)=\sum_j n_{j\vec{k}\sigma}\left|\int d^3r\,
  \Phi_{j\vec{k}\sigma}(\vec{r},t)\Phi^{\ast}_{i\vec{k}\sigma}(\vec{r},0)\right|^2
\end{equation}
These are associated with
the energies corresponding to the respective orbitals in the
ground state and integrated over the Brillouin zone (BZ), obtaining the expression:
\begin{equation}
  D_{\sigma}(E,t)=\sum_{i}\,\int_{\rm BZ}\,d^3k\,\delta(E-\epsilon_{i\vec{k}\sigma})\,g_{i\vec{k}\sigma}(t)
\end{equation}
For the following discussion it is important to keep in mind, that $D_{\sigma}(E,t)$ is used here as an effective tool to
visualize the change in the occupation of particular orbitals, differences between
features in $D_{\sigma}(E,t)$ do not necessarily correspond to real excitation energies.

The frequency-dependent dielectric function was
calculated in the framework of the random phase approximation
(RPA) in the limit $q\to 0$ neglecting microscopic contributions.
We compared these results with LR-TDDFT,
taking into account
many-body effects with the ``bootstrap'' (BS)
exchange correlation kernel,\cite{cn:Sharma11}
which includes self-consistently optimized long-range corrections.
These calculations were carried out with a
$16\times 16\times 6$ $k$-mesh for Fe$_1$/(MgO)$_3$(001),
$32^3$ $k$-points for bulk Fe and $48^3$ for bulk MgO.
% Appropriate broadening of the spectra was achieved by
% Fermi smearing with a width of 0.02\,Ry (0.272\,eV).

\section{Results}
\subsection{Geometry and electronic structure}\label{sec:static}
\begin{figure}
\centering
\includegraphics[width=0.8\columnwidth]{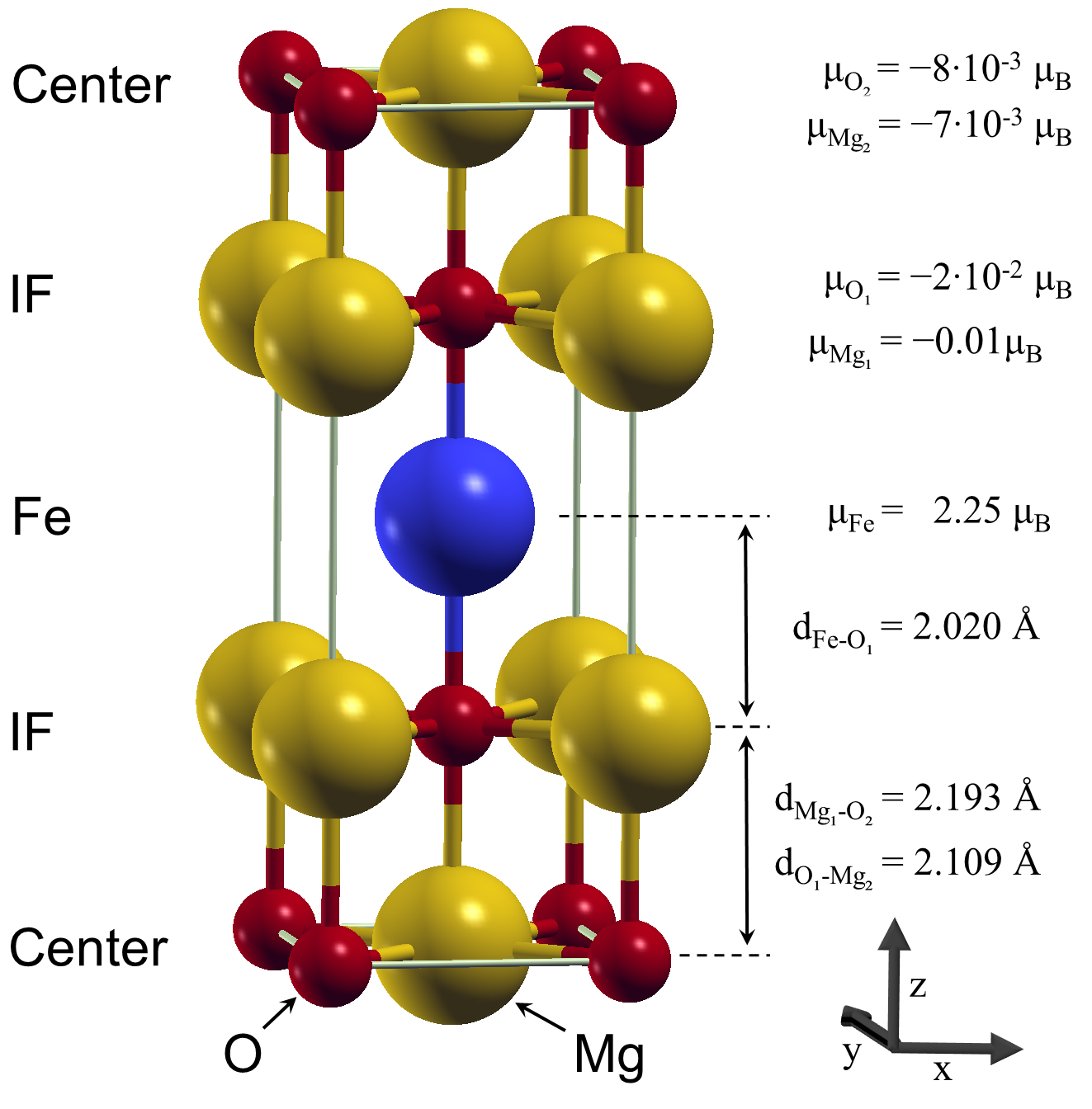}%
\caption{Unit cell of the four-layer Fe$_1$/(MgO)$_3$(001) heterostructure, consisting
  of one Fe-layer, the central MgO and one MgO interface layer (IF)
  after structural  optimization ($a$$\,=\,$$b$$\,=\,$2.972\,\AA{},$c$$\,=\,$8.432\,\AA{}).
  The numbers indicate the magnetic moments
  in the muffin-tin spheres obtained from Elk (LDA).
  \label{fig:cell}}
\end{figure}
Most of the studies of Fe/MgO heterostructures use
sharp interfaces,\cite{cn:Li91,cn:Butler01,cn:Feng09} while
the possibility of a FeO layer formation at the interface
has been addressed.\cite{cn:Meyerheim01,cn:Meyerheim02,cn:Luches05,cn:Valeri07,cn:Colonna09}
In our approach we adopted the first model, having O located apically to the
Fe,\cite{cn:Li91,cn:Butler01,cn:Feng09} as this configuration shows the lowest interface energy.\cite{cn:Cuadrado14} 
The superlattice consists of one Fe layer and  three layers of
bulk MgO.  The optimized cell parameters for the structure shown in Fig.\ \ref{fig:cell}
are $c$$\,=\,$8.259\,\AA{} and $a$$\,=\,$$b$$\,=\,$2.954\,\AA{}, indicating a
reasonable lattice  mismatch of  $-3.1\,\% $ % % $-1.1\,\% $
for MgO and $+4.4\,\%$ for Fe,
compared to the lattice constants of the respective bulk systems
obtained with similar technical settings.
%Revision B.3b, A.9
The distance between interface (IF) O and Fe is 2.020\,\AA, while the
MgO (IF) layer turns out to be slightly corrugated: the distance between the central
Mg and O (IF) amounts to 2.109\AA, whereas the one between Mg (IF) and central O is slightly larger, 2.193\,\AA, in agreement with earlier studies.\cite{cn:Meyerheim01,cn:Yu06,cn:Beltran12}
Using Elk for the optimization of the atomic positions we obtain
a very similar result.\footnote{Using Elk, Mg (IF) and O (IF) move slightly closer to the Fe plane.
The variation of the z-component difference to the VASP (PBE) result is
0.009\,\AA{} for Fe-Mg and 0.010\,\AA{} for Fe-O within Elk-PBE and
0.008\,\AA{} for Fe-Mg and 0.030\,\AA{} for Fe-O within Elk-LDA,
retaining the corrugation of the IF layer for both codes and functionals.}
The magnetic moment of Fe is 2.25\,$\mu_{\rm B}$,
close to the value for bulk bcc-Fe. On the other hand, thicker Fe films in Fe/MgO(001) exhibit enhanced magnetic moments (not shown here), consistent with previous first-principles theory\cite{cn:Ozeki07,cn:Yu06,cn:Bose16} and
experiment.\cite{cn:Sicot03,cn:Miyokawa05,cn:Jal15}
%Revision B.1

\begin{figure}
\centering
\includegraphics[width=\columnwidth]{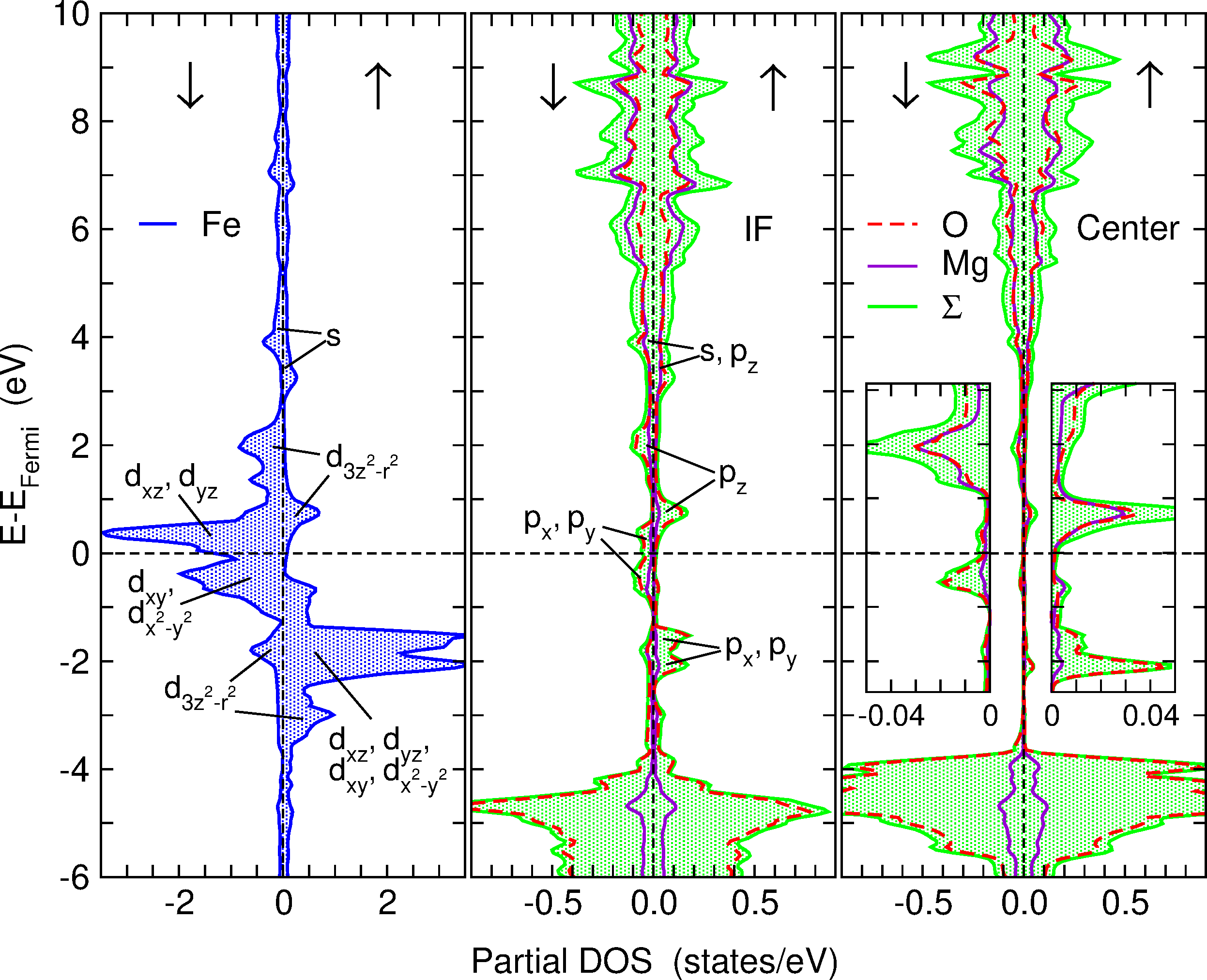}%
\caption{Spin-resolved layer-resolved electronic partial density of states (PDOS) of
  Fe$_1$(MgO)$_3$(001) obtained with Elk (LDA). The left panel shows the Fe-projected
  PDOS (blue lines), the center and right panel the PDOS from Mg, O and their sum ($\Sigma$) in the
  IF- and center-layer, respectively. Here, red dashed lines correspond to the O-projected
  PDOS while solid violet lines mark the contribution from Mg.
  Their sum is indicated by the green lines and background.
  For Fe and O (IF), additional labels indicate the orbitals which predominantly contribute
  to selected important features.
  \label{fig:PDOS}}
\end{figure}
Owing to the reduced dimensions the bandwidth of the single Fe layer is significantly narrowed compared to  bulk Fe  (cf. projected DOS in Fig.\ \ref{fig:PDOS}, left panel).
The Fermi-level is now pinned in a dip in the minority spin DOS
between the predominantly occupied
$d_{xy}$ and $d_{x^2-y^2}$ orbitals around $-0.5$\,eV and the
unoccupied $d_{xz}$ and $d_{yz}$ at $+0.3\,$eV.
The hybridization of Fe with the apical O leads to broadening and splitting of the majority spin $d_{3z^2-r^2}$ band into bonding and antibonding contributions centered at $-3$ and $+1\,$eV, respectively. The pseudo-gap
in the majority channel close to the Fermi level and the finite DOS in the minority spin channel imply a significant spin-dependence of the transport properties for the Fe$_1$/(MgO)$_3$(001) heterostructure.
The correspondence of the peaks in the Fe- and MgO-PDOS in
the adjacent interface (IF) layer (middle panel of Fig.\ \ref{fig:PDOS}) indicates strong hybridization between
Fe and O states 
in the energy range between -3\,eV and +3\,eV.
The sharp peaks in the Fe PDOS
around -2eV in the majority channel and around -0.4\,eV and +0.3\,eV in
the minority channel, which result from $d$-orbitals with in-plane orientation, 
coincide with the positions of $p_x$ and $p_y$ states of O (IF).
Likewise, we see marked features with $p_z$ character at +0.8eV and +2eV
above the Fermi level, which are the result from the strong hybridization with the
out-of-plane $d_{3z^2-r^2}$ orbitals of Fe.
The features at +0.3\,eV,  +0.8\,eV and +2\,eV
turn out to be of particular importance for the optical excitation discussed in in Sec.\ \ref{sec:dynamics}.
The energy range below -3\,eV and above +3eV in particular in the central MgO layer (the right panel of Fig.\ \ref{fig:PDOS}) resembles bulk MgO.
Still, some hybridization effects, %in particular,
for instance, the $p_z$-features above the Fermi level,
are visible in the central layer, indicating that the interface region extends to deeper MgO layers.
Nevertheless, the correspondence to the DOS of bulk MgO enables
us to estimate the band alignment at the Fe/MgO interface
within the % (semi-)local
local density and generalized gradient approximations used here.
We note that LDA and GGA exchange correlation functionals 
severely underestimate the band gap of MgO, which can affect the band alignment:
% Revision: A4
Using LDA with the same technical parameters as in our TDDFT setup
at the experimental bulk lattice constant the direct MgO gap at $\Gamma$ amounts to $4.64\,$eV,
compared to the experimental value of 7.7\,eV.\cite{cn:Williams67,cn:Roessler67,cn:Whited73}
The size of the band gap can be
corrected by using hybrid functionals or many body perturbation theory within the
GW approximation (e.g., Refs.\ \onlinecite{cn:Yamasaki02,cn:Schleife09,cn:Schimka10}) and BSE, that are beyond the scope of the present investigation.

% Revision B.2
It is not straightforward to identify the position of the bulk MgO gap in the heterostructure.
Interface states, as discussed above, arise from the hybridization with orbitals originating
from the metal layers. These decay exponentially in the MgO layers, as, e.g., shown in
Ref.\ \onlinecite{cn:Kang08} for a Si/SiO$_2$ interface.
In addition one may also consider a change of the gap due to
quantum confinement effects. Epitaxial strain arising from the
combination of different materials also affects the band structure.
Naively, one could infer from the right panel of Fig.\ \ref{fig:PDOS}, which shows
the valence band maximum in the central layer about $3.7\,$eV below the Fermi level,
that the conduction band minimum of MgO should be
located between $1.0\,$eV and $1.5\,$eV above $E_{\rm F}$. Consequently the $p_{x}$, $p_{y}$ and
$p_{z}$ interface states of O (IF) just above $E_{\rm F}$ would essentially lie within the gap,
whereas the second $p_z$ interface state at $+2.0\,$eV would already be within the conduction band.
A projection of the bulk bandstucture of MgO calculated in the same
Brillouin zone onto the bands of the Fe$_1$/(MgO)$_3$(001) heterostructure
(not shown) indeed reveals
a good match with the valence band maximum of MgO at $\Gamma$ with the
Fe/MgO-bands around -3.72\,eV.
Above the Fermi level, bands with significant Mg and O character
at $\Gamma$ at about 1.20\,eV and 2.15\,eV can be associated with the lowest conduction
bands of bulk MgO.
Naturally, the band alignment depends also on the thickness
of the slabs.
For comparison, in the Fe$_8$/(MgO)$_8$(001) heterostructure investigated in Ref. \onlinecite{cn:Rothenbach19}
the valence band maximum of the central layer is
located at -3.34\,eV, while states at $\Gamma$ which may be associated with the conduction band
minimum of the MgO slab are encountered at 1.67\,eV and 1.85\,eV.

\begin{figure}
\includegraphics[width=\columnwidth]{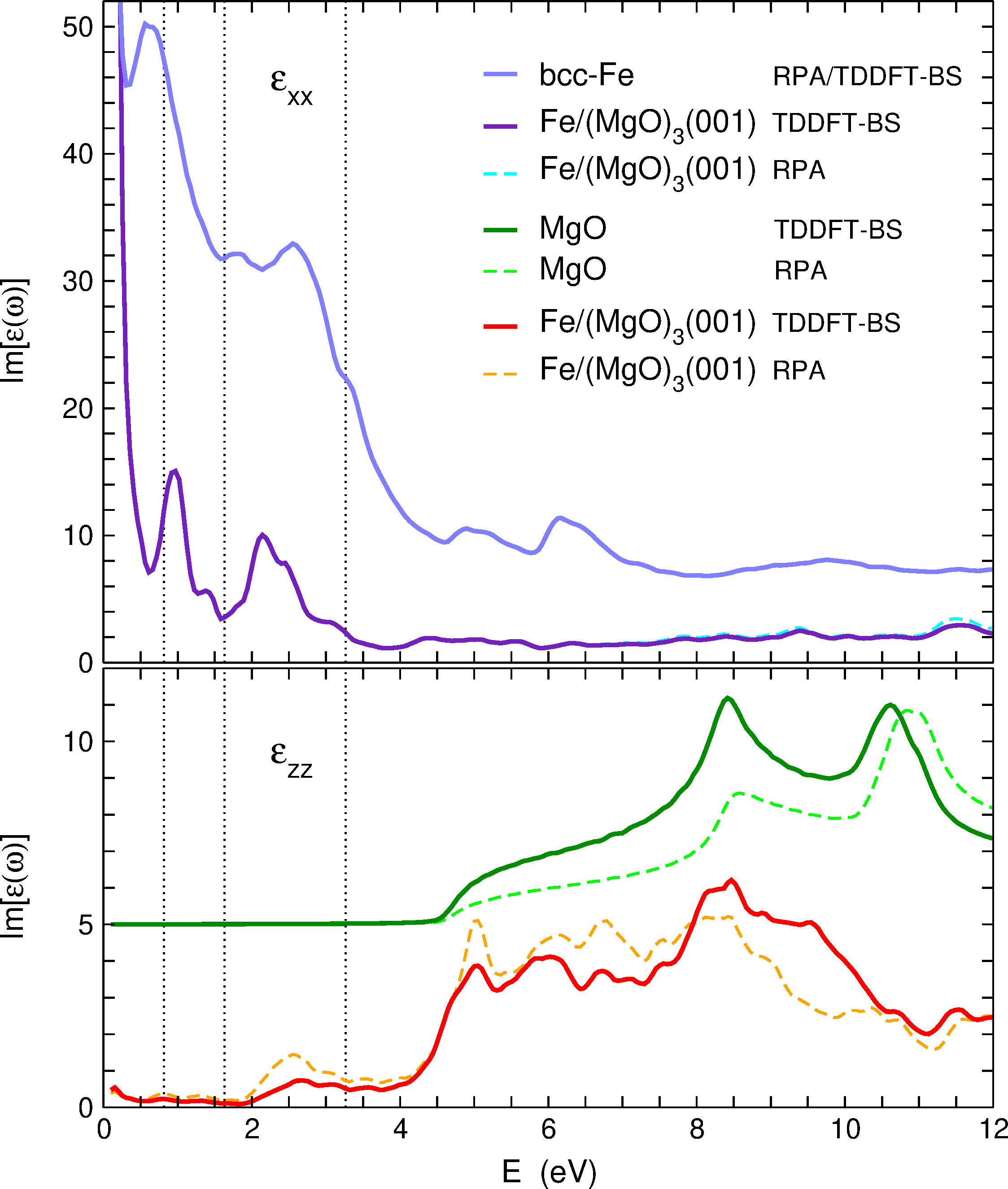}
\caption{Frequency dependence of the imaginary part of the dielectric tensor $\Im[\epsilon_{ij}(\omega)]$ in the
  Fe$_1$/(MgO)$_3$(001) heterostructure obtained with Elk within the random phase approximation (RPA) and LR-TDDFT, with
%  ALDA and
%Revision: A.9, B.3b
  the ``bootstrap'' (TDDFT-BS) exchange-correlation kernel.
  The upper panel shows the in-plane components
  $\Im[\epsilon_{\rm xx}(\omega)]$$\,=\,$$\Im[\epsilon_{\rm yy}(\omega)]$
  and the out-of-plane component $\Im[\epsilon_{\rm zz}(\omega)]$. The $\Im[\epsilon_{\rm xx}(\omega)]$ bears similarity
  with $\Im[\epsilon(\omega)]$ of bulk Fe, while  $\Im[\epsilon_{\rm zz}(\omega)]$ resembles bulk MgO. The spectra
  of bulk Fe and MgO are shifted vertically by a constant value of 5 for clarity.
  the vertical lines denote the laser frequencies used in our RT-TDDFT calculations.
  \label{fig:EPS}}
\end{figure}

\subsection{Absorption spectra from linear-response TDDFT}\label{sec:LRTDDFT}
% Revision: A.1
The optical properties of the heterostructure
are reflected in the
frequency dependent dielectric tensor $\epsilon(\omega)$, which can be derived from the
electronic structure in terms of the frequency-dependent Kohn-Sham susceptibility.
We calculated the imaginary part $\Im[\epsilon(\omega)]$, which describes the absorption of the material,
within the random phase approximation (RPA) and also
with TDDFT in the linear response scheme.
We compare $\Im[\epsilon(\omega)]$ of  the Fe$_1$/(MgO)$_3$(001) heterostructure with the
respective bulk systems,
i.\,e., bcc-Fe and rocksalt MgO, see Fig.\ \ref{fig:EPS}.
The bulk results agree well with the results of previous
calculations\cite{cn:Hamrlova16,cn:Botti04,cn:Schleife06,cn:Byun17}
obtained within the same level of approximation. For metallic Fe, including the
microscopic contributions from the exchange correlation kernel does not notably
alter the results, even with long-range corrections.
% Revision: B.3b
In contrast,
the ``bootstrap'' kernel (BS) leads to a significant renormalization of the optical absorption
spectrum for the insulating bulk MgO, which strongly improves the
agreement with experiment at larger excitation energies, 
whereas the appropriate description of the strong excitonic peak at 7.7\,eV
requires many-body perturbation theory
involving the Bethe-Salpeter equation.\cite{cn:Botti04,cn:Schleife09,cn:Byun17}

Due to the broken symmetry in $z$-direction,
the difference between the in-plane ($\epsilon_{\rm xx}$ and $\epsilon_{\rm yy}$, upper panel) and out-of-plane
($\epsilon_{\rm zz}$, lower panel) components of the dielectric tensor 
indicates a substantial dependence on the polarization of the incident light wave.
The in-plane components of the imaginary part of the dielectric tensor resemble (apart from a factor two in
magnitude) the respective quantity of bcc-Fe, with a large absorption in the low-energy region.
This corresponds to the setup where the incident light wave arrives perpendicular to the absorbing Fe-layers.
In contrast, the frequency dependence of the out-of-plane component $\Im[\epsilon_{\rm zz}(\omega)]$ bears similarity
to $\Im[\epsilon(\omega)]$ of bulk MgO, where absorption only takes place for photon energies above the gap.
In the  Fe$_1$/(MgO)$_3$(001) heterostructure,
the band alignment of MgO with metallic Fe and the presence of IF-states below and above $E_{\rm F}$
arising from hybridization causes weak absorption in the entire energy range below $4.7\,$eV.

Taking into account many-body effects in the framework of the bootstrap exchange-correlation kernel\cite{cn:Sharma11}
apparently does not make a substantial difference for metallic Fe and accordingly also the
in-plane components in our Fe$_1$/(MgO)$_3$(001) heterostructure.
For bulk MgO, the bootstrap kernel causes
a downward shift and rearrangement of the features at 8.5\,eV and 11\,eV as
compared to
RPA.
For the out-of-plane $\Im[\epsilon_{\rm zz}(\omega)]$ Fe$_1$/(MgO)$_3$(001), we rather notice
a redistribution of spectral weight in opposite direction for the TDDFT calculations %(ALDA and BS)
compared to the RPA.

\subsection{Real-time evolution of optical excitations}\label{sec:dynamics}
\begin{figure*}
\makebox[\columnwidth]{{\large\sf a)}\hfill$\hbar\omega$$\,=\,$0.816eV\hfill ~}%
\makebox[\columnwidth]{{\large\sf b)}\hfill$\hbar\omega$$\,=\,$1.632eV\hfill ~}\\[1ex]
\includegraphics[width=\columnwidth]{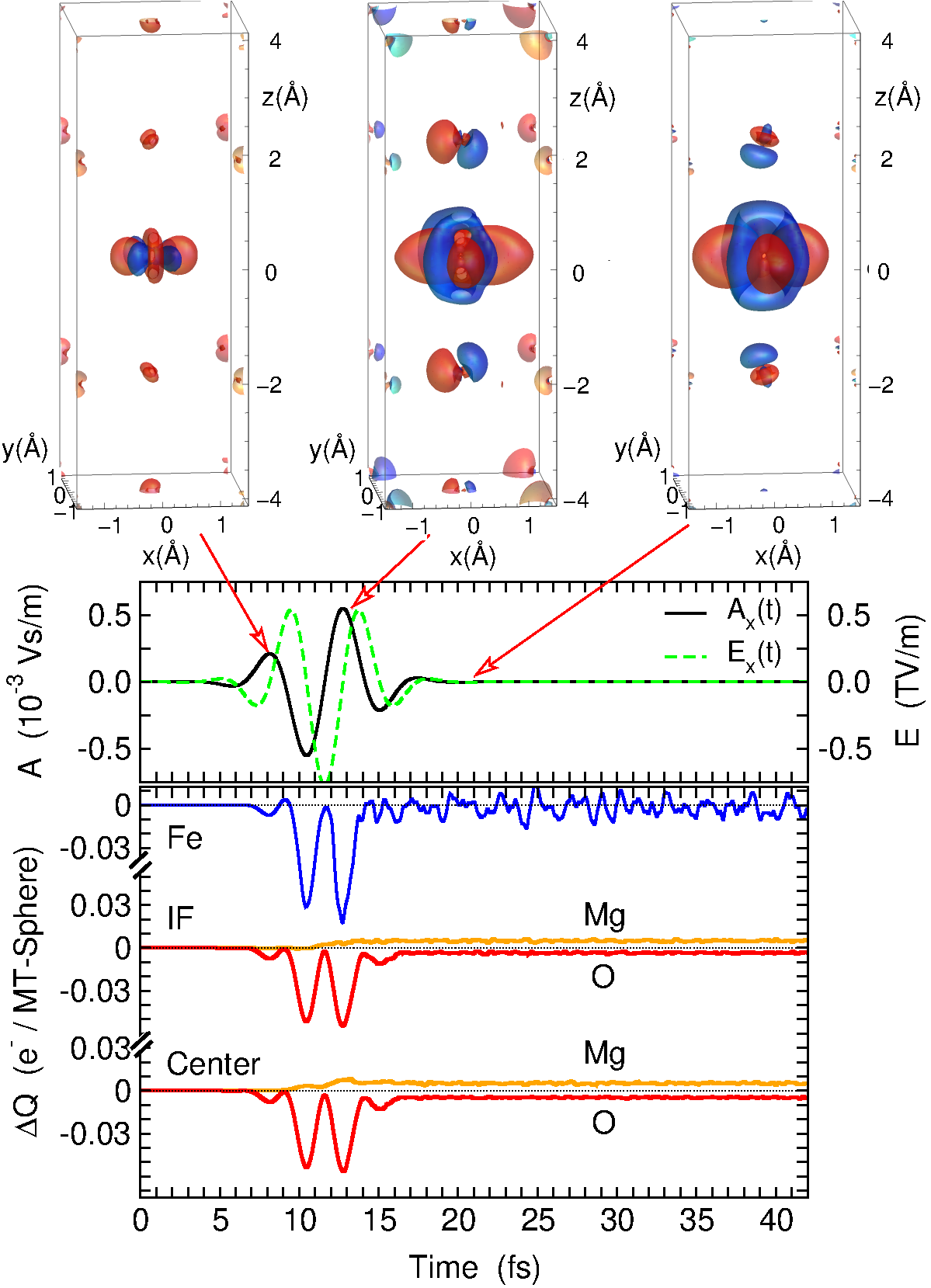}\hfill%
\includegraphics[width=\columnwidth]{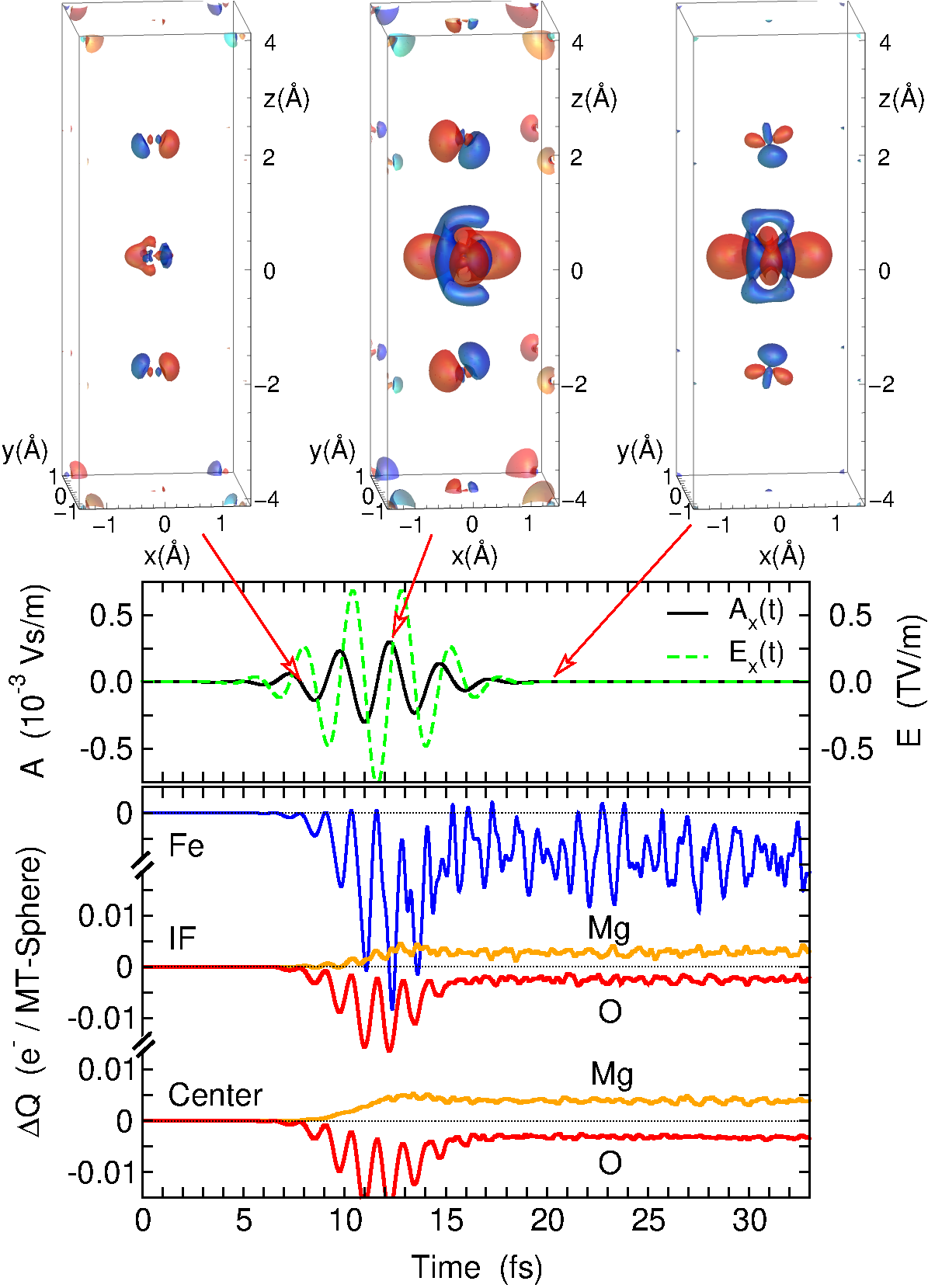}%
\caption{Time evolution of the charge distribution during a laser pulse
  with frequency (a)
  $\hbar\omega$$\,=\,$0.816\,eV and (b) $\hbar\omega$$\,=\,$1.632\,eV.
  The upper panels show the change in electronic density $\rho(\vec{r},t)$
  relative to the initial state $\rho(\vec{r},t)-\rho(\vec{r},0)$ in terms of isosurfaces corresponding
  to charge levels of $\pm 2\times 10^{-3}\,e_0/a_{\rm B}^3$   at three different points in time
  ($t$$\,=\,$8.1\,fs, 11.8\,fs and 20\,fs).
  Red color denotes a depletion of
  electronic (negative) charge, blue an accumulation inside the isosurface.
  The panels in the central row represent the time-dependence of the external
  vector field $\vec{A}(t)$$\,=\,$$A_x\,\vec{e}_x$
  (solid black line) and the resulting electric field
  $\vec{E}(t)$$\,=\,$$-\partial \vec{A}/\partial t$
  (dashed green line).
  The lower panels depict the time-dependent change in the electronic charge
  $\Delta Q$$\,=\,$$Q(t)-Q(0)$ residing inside
  the muffin-tin-spheres of Fe, Mg and O in the different layers.
  \label{fig:CHG}}
\end{figure*}
\begin{figure*}
\includegraphics[width=1.5\columnwidth]{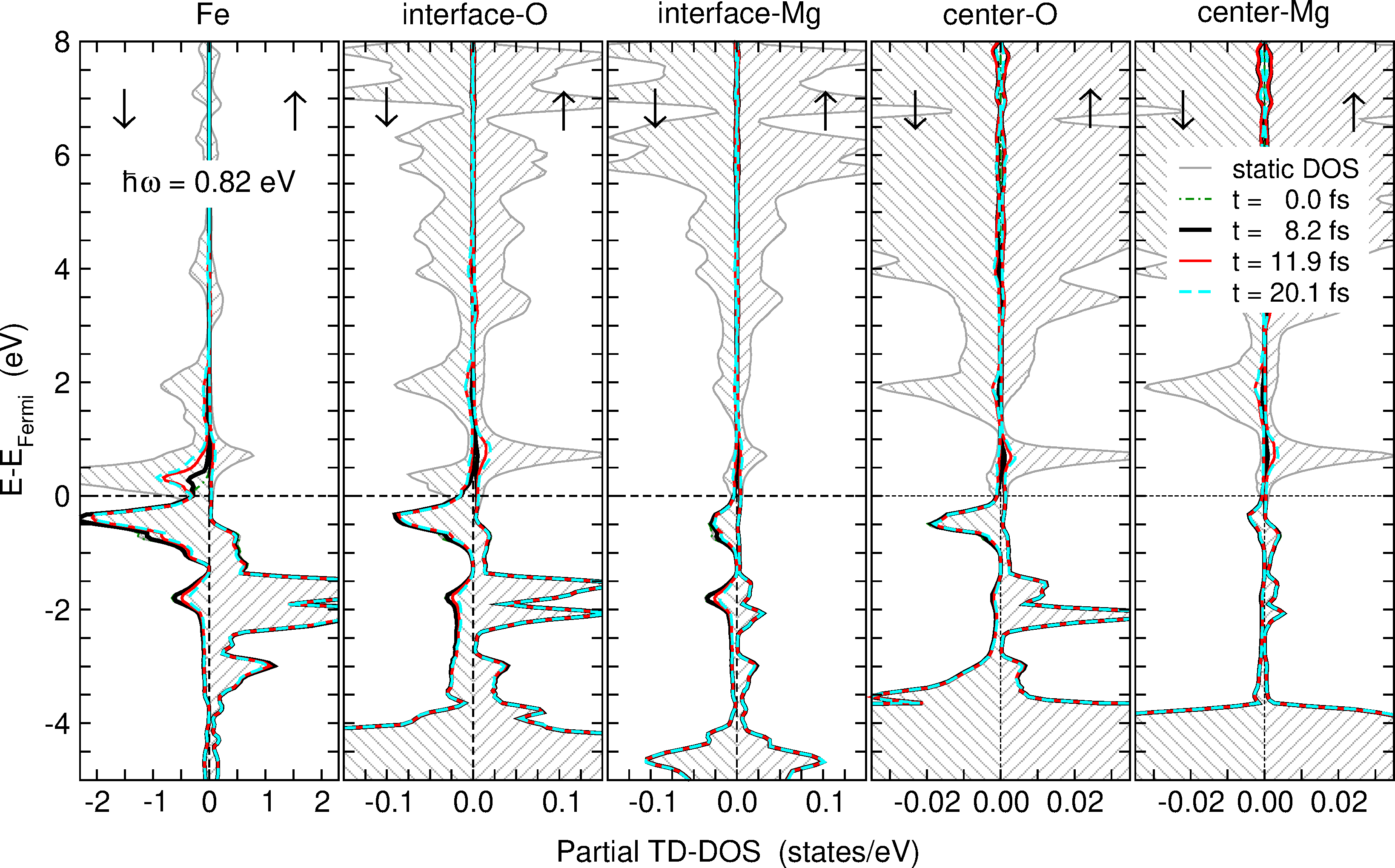}% {figs/TDDOS-Fe1MgO3-td03-highk-new.png}%
\caption{Site-resolved time-dependent partial spectral function
% Revision B.3d
%  (density of states)
  $D_{\sigma}(E,t)$
  for $\hbar\omega$$\,=\,$0.816eV evaluated
  at four different points in time, corresponding to the snapshots in the
  charge density in Fig.\ \protect\ref{fig:CHG}a: $t$$\,=\,$$0$ (thin dash-dotted line),
  $t$$\,=\,$$8.2\,$fs (thick black line), $t$$\,=\,$$11.9\,$fs (thick red line),
  $t$$\,=\,$$20.1\,$fs (thick dashed cyan line).
  The gray hatched areas in the background shows the corresponding static partial DOS,
  obtained from a ground-state calculation with the same technical settings.
  \label{fig:TDPDOS}}
\end{figure*}
\begin{figure*}
\includegraphics[width=\textwidth]{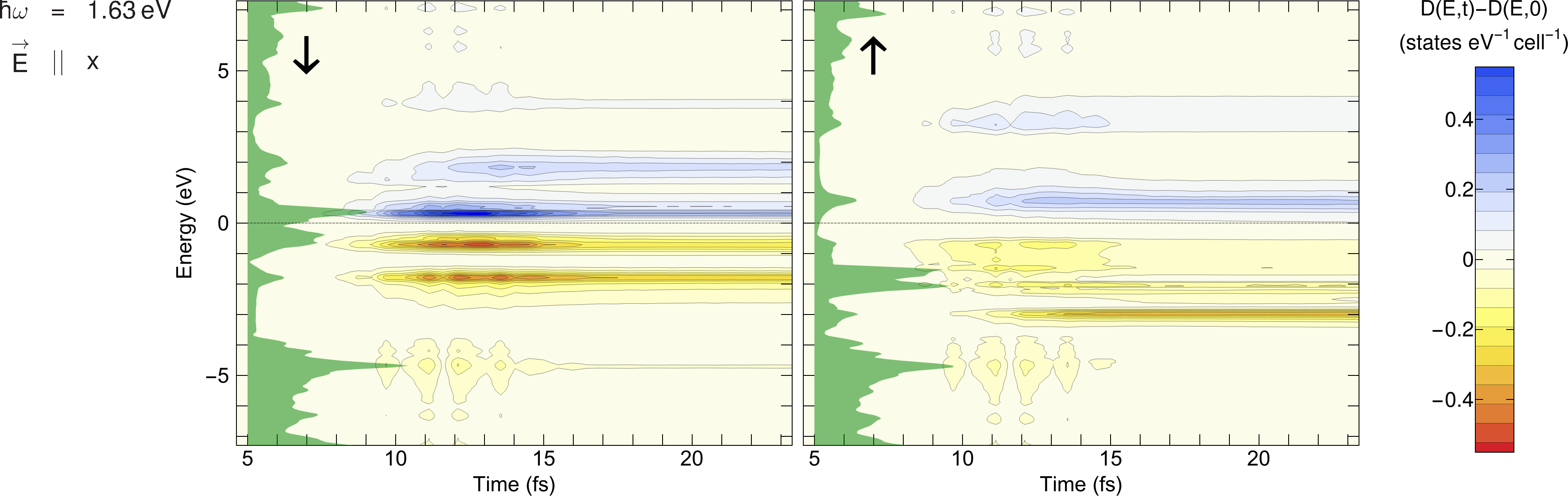}% {figs/leg_td06x_total.png}
\caption{Energy and time-resolved changes in the $\Delta$TDDOS,
  $\Delta D_{\sigma}(E,t)=D_{\sigma}(E,t)-D_{\sigma}(E,0)$ for the
  laser pulse shown in Fig.\ \protect\ref{fig:CHG}b ($\hbar\omega$$\,=\,$1.632eV).
  The left panel refers to the minority spin channel,
  the right panel to the majority spin channel. Orange and red colors refer to a depletion of occupation,  blue colors to an increase. The green area at the left edge of each panel indicates the shape of the 
  spin-resolved total density of states.
  \label{fig:TDOS}}
\end{figure*}
To investigate the dynamics of excitation and charge transfer into and across the interface, we
chose three different frequencies for optical excitations,
$\hbar\omega$$\,=\,$$0.816\,$eV,
$\hbar\omega$$\,=\,$$1.632\,$eV and $\hbar\omega$$\,=\,$$3.265\,$eV, while keeping
the laser fluence constant.
All three excitations are clearly within the gap of bulk MgO,
thus a direct excitation from the valence band to the conduction band cannot take place. However, the band alignment in our heterostructure suggests that the conduction band edge of MgO is
located between $1.2$ and $2.2\,$eV above the Fermi level,
as discussed in Sec.\ \ref{sec:static}. Thus the lower frequency produces
excitations in Fe with final states predominantly
below the conduction band edge of MgO, while with $\hbar\omega$$\,=\,$$1.632\,$eV one can already
address states in the central layer which correspond to the conduction band of MgO.
% Revision A.5
For completeness, we considered a third energy, $\hbar\omega$$\,=\,$$3.265\,$eV,
which clearly reaches Fe-states above the conduction band minimum of MgO,
while a direct excitation of carriers between states
corresponding to the MgO valence and conduction band is still inhibited.
Besides the frequency dependence, we explore the effect of different polarization directions
of the incoming light wave.

\subsubsection{In-plane polarization}
\begin{figure*}
\includegraphics[width=\textwidth]{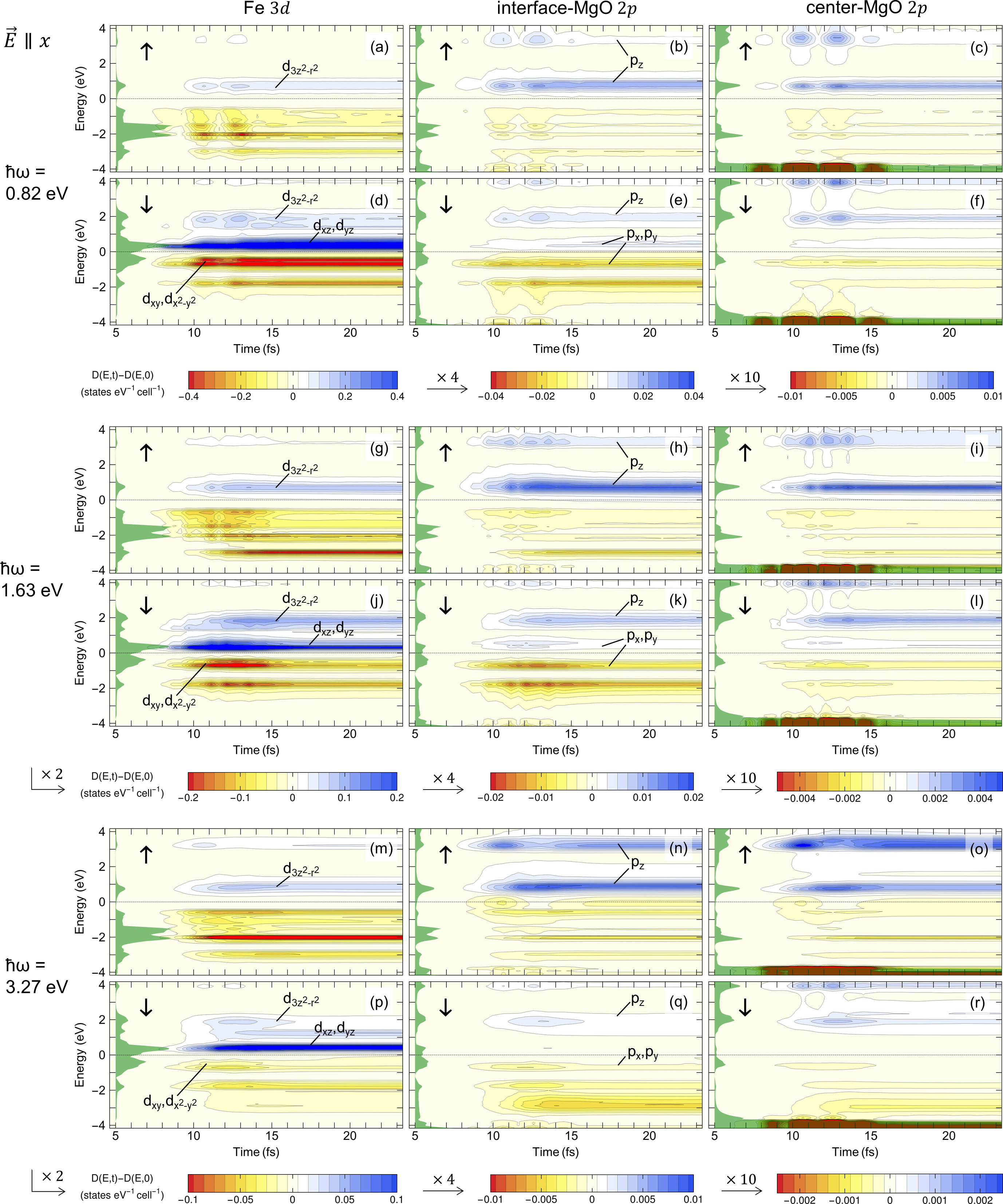}% {figs/TDDOS-all.png}%
\caption{Energy and time-resolved changes in the atom- and orbital-resolved  $\Delta$TDDOS
  for the laser frequencies (a)-(d) $\hbar\omega$$\,=\,$0.816\,eV,
   (g)-(l) $\hbar\omega$$\,=\,$1.632\,eV and (m)-(r)  $\hbar\omega$$\,=\,$3.265\,eV. 
  The left column refers to the Fe-$3d$ states, the middle column to the IF-MgO-$2p$ states
  and the right column to the center-MgO-$2p$ states.
  Note that the scale of  the color coding (and the static partial DOS indicated by the
  green area) changes between the panels.
  \label{fig:PTDOS}}
\end{figure*}
\begin{figure}
\includegraphics[width=\columnwidth]{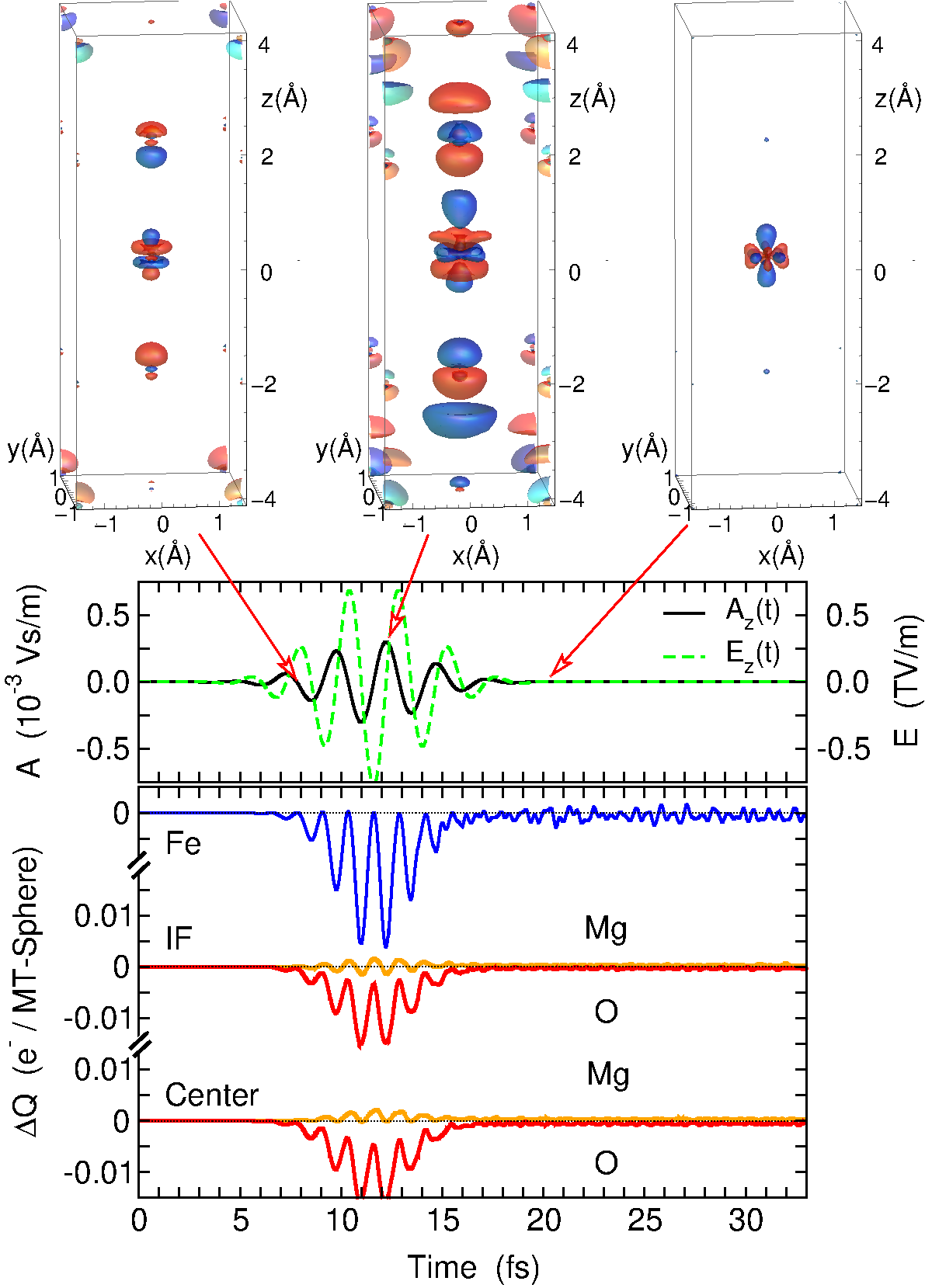}% {figs/CHG-td06-para.png}%
\caption{Time evolution of the charge distribution during a laser pulse
  with frequency $\hbar\omega$$\,=\,$1.632\,eV with out-of-plane polarization of the
  vector field $\vec{A}(t)$$\,=\,$$A_z\,\vec{e}_z$. Same colors and lines as in
  Fig.\ \ref{fig:CHG}.
  \label{fig:CHG06z}}
\end{figure}
\begin{figure*}
\includegraphics[width=\textwidth]{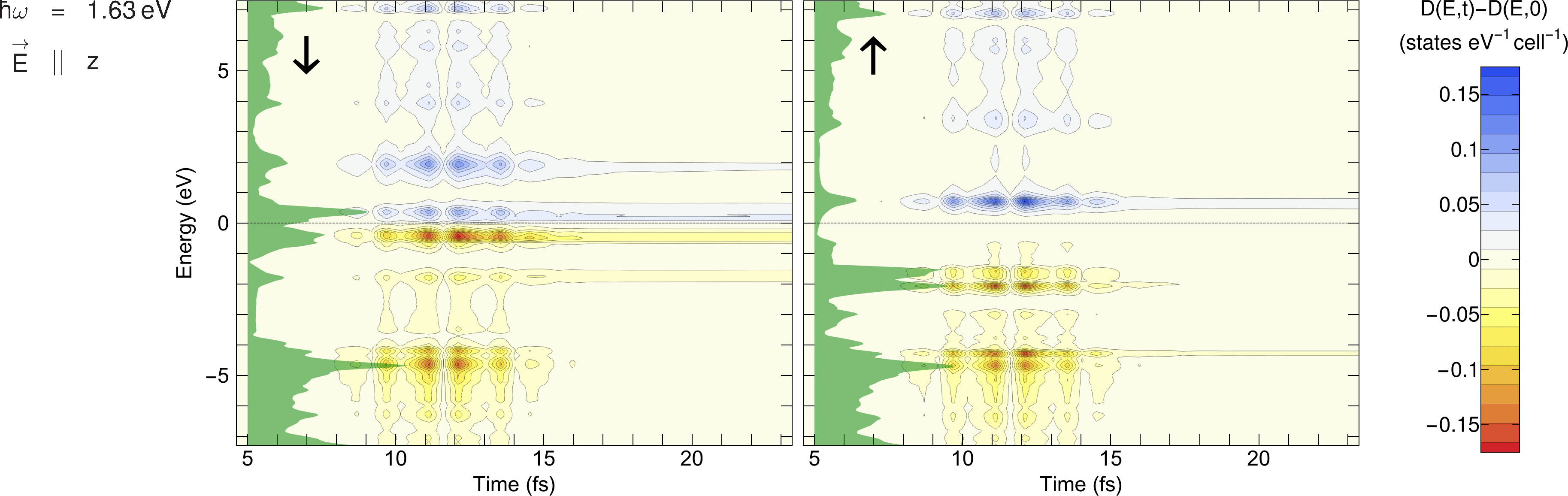}% {figs/leg_td06z_total.png}
\caption{Energy and time-resolved changes in the total  $\Delta$TDDOS
  $D_{\sigma}(E,t)-D_{\sigma}(E,0)$ for the
  laser pulse shown in Fig.\ \protect\ref{fig:CHG06z}. Same colors and symbols as in
  Fig.\ \ref{fig:TDOS}.
  \label{fig:TDOS06z}}
\end{figure*}

In a typical setup in pump-probe experiments the
laser pulse propagates normal to the sample surface.
The electrical field components are then confined within the $x$-$y$-plane.
Figure \ref{fig:CHG}a presents the evolution of the charge distribution
during and after an in-plane polarized laser pulse of $\hbar\omega$$\,=\,0.816\,$eV.
While this energy is not yet sufficient to reach the MgO conduction band in the heterostructure,
Figure \ref{fig:CHG}b shows the corresponding results for a pulse with
$\hbar\omega$$\,=\,1.632\,$eV, which could allow
for the transfer of an excitation into states of the central
layer, which correspond to the conduction band minimum of MgO.
% Revision: B.3c
Apart from the smaller amplitude of the charge fluctuation
(cf. the lower panel of Figure \ref{fig:CHG}b), the electric field causes a 
rather analogous response in both cases. The upper panels visualize
snapshots of the electron density redistribution at
$t$$\,=\,$$8.2\,$fs, $t$$\,=\,$$11.9\,$fs, $t$$\,=\,$$20.1\,$fs, 
which correspond to the onset of the pulse, the largest amplitude of the vectorpotential and
after the decay of laser pulse, respectively.
Our results show that
the redistribution is strongest around the Fe sites but affects the whole MgO region. 
At the onset of the pulse, charge dipoles form in particular at Fe and  the apical O sites
and to a lesser extent also at the Mg and central O sites. At the Fe site, the redistribution
of charge exhibits a strong orbital dependence: charge is depleted  essentially from the  in-plane
$d$ orbitals ($d_{xy}$, $d_{x^2-y^2}$)  and accumulated in out-of-plane orbitals ($d_{3z^2-r^2}$,  $d_{xz}$,  $d_{yz}$).
Similarly, an accumulation is observed in the $p_z$ orbital of the apical oxygen. 
The oscillation of the electrical field in $x$-direction
forces the dipoles around the apical O to rotate with the
accumulated (blue) electron bulb always located in between the O and Fe sites.
With decreasing field strength the redistribution of charge largely regresses,
in particular in the central layer and at Mg (IF).
The prominent features at the Fe and the apical O (IF) sites remain, finally reaching a steady,
weakly oscillating state (an animation of the time-dependent evolution of the charge
distribution is provided as supplemental material \cite{cn:supplement}).

The integrated charge contained in the muffin-tin spheres centered
around each lattice site shown in the lower panels of Fig.\ \ref{fig:CHG} renders
a complementary view on the dynamical evolution of the charge distribution.
During the pulse, the electric field accelerates the electronic charge which temporarily
leaves the muffin-tin spheres. When the field reaches its negative turning point, the charge
is largely restored.
This picture is only encountered for the Fe and O-spheres, which harbor a significant amount of valence
charge as opposed to Mg. The reduction of charge at Fe and O sites starts at approximately $t$$\,=\,$7\,fs,
while a much smaller accumulation in the Mg spheres is encountered later, at $t$$\,=\,$9\,fs. % , which further.
Since the total charge in the system is conserved, we conclude that with each oscillation of the field,
charge subsequently swashes
out of the O- and Fe-spheres into the interstitial, from where a fraction eventually
moves into the positively charged Mg-spheres.
Thus, the light field effectively pumps electronic charge from the O towards the Mg sites.
Interestingly, the behavior of the IF and central layer is qualitatively similar but differs
in magnitude, which
is a consequence of the central MgO layer of this thin heterostructure
yet reaching bulk optical properties,
This effect remains stationary even after the laser pulse has been switched off.

Further insight into the evolution of the electronic structure can be obtained from the
spin- and site-resolved time-dependent spectral function $D_{\sigma}(E,t)$.
This is shown in Fig.\ \ref{fig:TDPDOS}
for $\hbar\omega$$\,=\,$0.816\,eV for distinct times $t$.
At $t$$\,=\,$0 all states below $E_{\rm F}$ are occupied,
apart from the thermal broadening. With time evolving, excitations build
up in particular in the minority spin channel of Fe, 0.3\,eV above the Fermi level,
obviously fostered by the availability of $d_{xz}$ and $d_{yz}$ states.
A similar trend can be observed at the $d_{3z^2-r^2}$ peak at 0.75\,eV
in the majority channel. Since this peak is much smaller, the effects are not as
pronounced and barely visible in Fig.\ \ref{fig:TDPDOS}. However, they become apparent,
if we look on a magnified scale at the corresponding hybridized IF-states
of the apical O atoms.
Compared to O (IF), we encounter a significantly smaller response from
the states above the Fermi level for Mg (IF). In the central layer
the change in occupation numbers is significantly smaller but still visible.
In contrast to the IF, the center-Mg response is rather similar to O.

For a detailed analysis of the temporal evolution of the electronic structure
after the onset of the laser pulse,
we use the time-dependent changes in the spectral weight
of the occupied states ($\Delta$TDDOS),
defined as
% Revision B.3d
the difference of the % density of states
time-resolved spectral function
$\Delta D_{\sigma}(E,t)$ at time $t$ with respect to the initial
state , i.\,e., $D_{\sigma}(E,t)-D_{\sigma}(E,0)$.
The contour plot of the total $\Delta$TDDOS in
Fig.\ \ref{fig:TDOS} for the case $\hbar\omega$$\,=\,$1.632\,eV exhibits distinct features, which
are directly correlated to regions of high density of states of Fe in the respective spin channel
(cf.\ Fig.\ \ref{fig:PDOS} and the green area in
each panel of Fig. \ref{fig:TDOS}).
Independent of the laser energy,
the minority channel exhibits the
largest $\Delta$TDDOS, as the peaks in
the ground state majority DOS are generally smaller in the vicinity of $E_{\rm F}$.
Here, the largest changes
occur during the pulse, afterwards we see an essentially stationary picture,
keeping in mind that
relaxation processes beyond the electronic  time scale are not accounted for in
the current RT-TDDFT modelling. 

We observe the fastest response to the laser pulse in the energy range closest
to the Fermi level, whereas the change in occupation appears retarded for the more distant states.  
We interpret this as a signature of a sequential excitation process, where
previously created holes
below $E_{\rm F}$ are refilled by electrons originating from a lower lying state.
Multiple subsequent excitations might also account for the rather strong depletion of the $d$-states close to
$-2\,$eV (minority channel) and $-3\,$eV (majority channel)
for $\hbar\omega$$\,=\,$1.632\,eV.

The isolated features, which appear
in particular at low and high energies (below $-4\,$eV and above $+4\,$eV) far beyond the excitation
energy of the laser correlate to the extrema of the vector potential. Here, the
electric field has its largest contribution to the generalized momentum of the electrons,
which directly impacts their kinetic energy. Therefore,
we ascribe these effects to a
partially non-linear response arising from the large amplitude of the laser field in combination with
the comparatively broad frequency distribution
in our Gaussian-shaped finite-length pulses.
The features decay quickly with decreasing vector potential,
but a small persistent change in occupation may remain, even after the pulse is over.

We now address the dynamic evolution of the site- and orbital-resolved
partial  $\Delta$TDDOS, % spectral function,
which we compare in Fig.\ \ref{fig:PTDOS} for the three laser
frequencies.
We selected the Fe-$3d$ and the sum of the  Mg-$2p$ and O-$2p$ orbitals
(the latter resolved for IF and center layers)
which provide the largest contributions for the respective layers.
The Fe-$d$ contribution
is the largest in absolute numbers and thus closely resembles the time
evolution of the total spectral function.
This confirms that the light predominately interacts with the $d$-electrons of the Fe-layer.
In accordance with our calculation of the frequency dependent dielectric tensor
(Fig.\ \ref{fig:EPS} in Sec.\ \ref{sec:LRTDDFT}) we find a considerably weaker interaction (absorption) of
the laser field with the electronic structure
for increasing laser frequencies (note the change in the scale of the contour plots
in Fig.\ \ref{fig:PTDOS}, where the distance between two contour lines is reduced
from the top to the bottom panels).

Concentrating on the IF-MgO (center panels in Fig.\ \ref{fig:PTDOS}),
we observe a particularly strong occupation of the O-$p_z$ states above the Fermi level
(+0.8\,eV in the majority channel and +2\,eV in the minority channel) which are present
at these energies due to the hybridization with the Fe-$d_{3z^2-r^2}$ states.
In relation to the excitation taking place in the Fe layer, these
contributions become considerably more pronounced with increasing laser frequency.
In turn, we find a much
weaker occupation of the O-$p_x$ and $p_y$ orbitals
above $E_{\rm F}$ as we would expect from the
size of the peaks in the static DOS in Fig.\ \ref{fig:PDOS}.
In contrast to the $p_z$ orbitals, their
$\Delta$TDDOS strongly decreases with energy and
nearly vanishes at $\hbar\omega$$\,=\,$3.265\,eV. 
Below $E_{\rm F}$, there is a considerable
depletion of the in-plane $p$-orbitals, which is again a consequence of the
hybridization with the in-plane Fe-$3d$ orbitals.
As for the IF layer, we find a comparatively large % time-dependent occupation
$\Delta$TDDOS of the states at +0.8\,eV (majority channel)
and +2\,eV (minority channel) in the central layer, which
now results from both, O and Mg (cf. Fig.\ \ref{fig:TDPDOS}).
In contrast, however, we do not observe a similar depletion of the states below $E_{\rm F}$.
% Revision: A.4
We find mainly additional occupation
of particular states above $E_{\rm F}$ without concomitant
de-occupation of corresponding states below, which we would expect in case of a direct excitation
within the layer.
Therefore, our observation might rather
be considered as a signature of excitations propagating into the interface and
we can infer that -- in contrast to the in-plane-orbitals --
hybridization between the out-of-plane orbitals
at the interface, i.\,e., Fe-$d_{3z^2-r^2}$ and O-$p_z$ are of particular importance for
the propagation of excitations into deeper MgO layers away from the interface.
%across multiple IF-layers.
Since the O-$p_z$ states in the minority channel are located above
the conduction band edge, the excitation may even propagate into bulk MgO.
This aspect needs further investigation, since it can be affected by
the underestimation of the MgO band gap  within LDA.

\subsubsection{Cross-plane polarized pulse}
A qualitatively different response is obtained,
when the polarization of the vector potential is in out-of-plane direction,
which we briefly discuss here for the frequency $\hbar\omega$$\,=\,$1.632\,eV.
As illustrated by Figs.\ \ref{fig:CHG06z} and \ref{fig:TDOS06z},
the charge clouds around the atoms now oscillate in vertical direction, which leads again
to a decrease of charge inside the muffin-tin-spheres of Fe and O, comparable to the respective in-plane case.
The time evolution of the charge density shows that persistent excitations remain
beyond 20\,fs, but
these are essentially located at the Fe site (Fig.\ \ref{fig:CHG06z}).
In contrast to the in-plane polarized pulse,
we do not encounter a significant persistent charge transfer
from O to Mg.
From the time-dependence of the
occupation numbers in Fig.\ \ref{fig:TDOS06z} we conclude that the charge-redistribution at the Fe-site is mainly
related to a redistribution of
states in the minority channel across the Fermi level, which occurs only in an interval of $\pm0.5\,$eV around
$E_{\rm F}$.
As in the in-plane case, we also see strong changes in $D_{\sigma}(E,t)$ at the maximum of the vector potential. Most of these
features vanish, however, when the pulse is over.
This means that a permanent absorption of photons
has not taken place in this case.
We infer from the absence of strong transient changes after the laser pulse
that the cross-plane electric field component of a
laser pulse arriving at a grazing angle will not contribute substantially
to the excitation process in and across the interface.
The distinct behavior for in- and out-of-plane polarization is consistent
with the anisotropic bahavior of the imaginary part of the dielectric tensor
discussed in Sec.\ \ref{sec:LRTDDFT}.

\section{Conclusions}\label{sec:conclusion}
We simulate the excitation of a metal-insulator heterostructure
by an ultrashort laser pulse  in the visible
and infrared frequency range with real-time TDDFT, revealing the
dynamic evolution of electronic excitations at the interface.
We consider an ultrathin Fe$_1$/(MgO)$_3$(001)
heterostructure and photon energies which are below the band gap of
bulk MgO, but sufficient for a local excitation in the Fe layer,
which could induce an electronic transfer into the  conduction band of MgO,
subsequently.
Furthermore, we take into account that the MgO layers at the interface
have states close to the Fermi-level, which jeopardize
the insulating character in cross-plane direction and will thus help to
propagate the excitation into the MgO. The orientation of the
orbitals at the interface is an important factor for the propagation
of excitations into the insulator.

We observe that excitations and their possible propagation
through the system take place nearly instantaneously, i.\,e., during the
laser pulse.
After the pulse, the system reaches an essentially stationary state,
which remains unaltered up to at least 42\,fs.
For normal incidence, 
when the electric field of the laser light is polarized within the Fe-plane,
we observe
particularly strong accumulations in states which arise from the hybridization
of the $p_z$-orbitals of O with the $d_{3z^2-r^2}$ orbitals of Fe at the interface.
In the central layer, which has no direct contact to Fe, we still find states
within the gap, which share this particular sensitivity to the light pulse.
We consider this as a strong indication that hybridization of orbitals, which are oriented
along $z$ (i.\,e.\ in cross-plane direction) is of particular importance to inject excitations
across the interface into deeper layers.
We further observe a weak charge transfer from O to Mg in the interface but also in
the central MgO layer.
For cross-plane polarization, strong effects are confined to the duration of the pulse,
the steady state involves mainly the Fe-layer.
Charge transfer from O to Mg and persistent laser-induced excitation of states above the
Fermi level are suppressed.

Using in-plane polarized light, a heterostructure with a favorably hybridized orbitals
might thus potentially be used for the optical injection of carriers with a specific
spin-polarization into the conduction band of a semiconductor or insulator,
whereas the out-of-plane
component does not contribute to this process substantially.
The dependence on the light polarization direction is understood based on the
frequency dependent dielectric tensor %optical properties 
calculated using LR-TDDFT, which involves a strong anisotropy:
The imaginary part of the $xx$-component implies significant absorption, resembling
$\Im[\epsilon_{xx}]$ of bulk Fe.
In turn, the $zz$-component rather
bears similarities to the optical response of MgO, but we still find a small
but finite absorption in $\Im[\epsilon_{zz}]$ in the low frequency range due to the interface states.

Our minimal model system gives a first comprehensive
overview on the fundamental response of a metal-insulator heterostructure
after a laser excitation, while
future work shall address more realistic heterostructures with thicker
slabs and many body effects,\cite{cn:Yamasaki02,cn:Schleife09,cn:Schimka10}
that are currently not available in RT-TDDFT.

\begin{acknowledgments}
  We gratefully acknowledge discussions
  with Sangeeta Sharma (MPI Berlin),
  Peter Elliot (MPI Halle), Andrea Eschenlohr,
  Uwe Bovensiepen, Heiko Wende and Klaus Sokolowski-Tinten
  (University of Duisburg-Essen).
  Funded by the Deutsche Forschungsgemeinschaft (DFG, German Research Foundation)
  -– project number 278162697 -– SFB 1242 (subproject C02).
  Calculations were carried out on the MagnitUDE supercomputer system (DFG grants no.\ INST 20876/209-1 FUGG and INST 20876/243-1 FUGG).
\end{acknowledgments}

%\bibliography{MgO,TDDFT,TDDFT2,TDDFT3,ufast1,FeMgO,FeMgO2,DFT,ufast2}
%

\end{document}